%% file: main.tex
\documentclass[10pt]{article}
\usepackage{a4}
\usepackage{ifpdf}
\ifpdf
\usepackage{hyperref}
\else
\usepackage{url}
\fi
\usepackage{rotating}
\usepackage{tabularx}
\usepackage{color}
\usepackage{graphicx}

\def\inred#1{\textcolor{red}{#1}}
\def\ingreen#1{\textcolor{green}{#1}}
\def\P{\mathcal{P}}
\newtheorem{definition}{Definition}
\newtheorem{theorem}{Theorem}

\begin{document}

\title{Solving Package Dependencies: \\
 From EDOS to Mancoosi\footnote{The research leading to these results has
   received funding from the European Community's Seventh Framework Programme
 (FP7/2007-2013) under grant agreement n${}^\circ$214898.}}
\author{Ralf Treinen and Stefano Zacchiroli \\
 Laboratoire Preuves, Programmes et Syst\`emes \\
 Universit\'e Paris Diderot, Paris, France \\
 \texttt{\{treinen,zack\}@\{pps.jussieu.fr,debian.org\}}}

\maketitle

\begin{abstract}
 Mancoosi (Managing the Complexity of the Open Source Infrastructure)
 is an ongoing research project funded by the European Union for
 addressing some of the challenges related to the ``upgrade problem''
 of interdependent software components of which Debian packages are
 prototypical examples.

 Mancoosi is the natural continuation of the EDOS project which has
 already contributed tools for distribution-wide quality assurance in
 Debian and other GNU/Linux distributions. The consortium behind the
 project consists of several European public and private research
 institutions as well as some commercial GNU/Linux distributions from
 Europe and South America. Debian is represented by a small group of
 Debian Developers who are working in the ranks of the involved
 universities to drive and integrate back achievements into Debian.

 This paper presents relevant results from EDOS in dependency
 management and gives an overview of the Mancoosi project and its
 objectives, with a particular focus on the prospective benefits for
 Debian.
\end{abstract}

\input{introduction}

\input{edos}

\input{mancoosi}

\bibliography{mancoosi}
\bibliographystyle{alpha}

\end{document}

%% file: introduction.tex
\section{Introduction}
\label{sec:introduction}

Building and maintaining a free software distribution is a challenging
task. A user expects to be able to install any selection of packages
from the distribution on his machine, and that the installation goes
smoothly and results in a working system with the desired
functionality. Any requirement, for instance the need of installing
certain auxiliary packages from the distribution, should be detected
by the tools coming with the distribution, and should be satisfied
automatically whatever packages the user wishes to
install. Incompatibilities in user wishes should be detected and
reported back to the user with a satisfying explanation. Software is
expected to be readily available in its latest version, of course
well-tested without any bugs or any remaining incompatibilities with
other software components.  All this is expected to work smoothly on a
wide range of architectures and system configurations. 

It is the task of a package maintainer to do her best to satisfy these
expectations. Luckily, a maintainer has at her disposition a
sophisticated infrastructure, a knowledge base of policies and best
practices, and the support of her fellow developers. On the other hand
the maintainer is also faced with upstream authors who usually have
their own ideas about how their software is supposed to be compiled,
or how it should interact with the rest of the system.

The EDOS research project (for \emph{Environment for the development and
Distribution of Open Source software}) had the objective of coming
to help and to provide FOSS distributions with better tools to help
them do their job. The project was funded by the European Commission
under the IST (\emph{Information Society Technologies}) activities of
the 6th Framework Programme. Besides several public research
institutions from different European countries and some small
enterprises in the FOSS business there were two commercial GNU/Linux
distributions in the project: Mandriva from France who is building one
of the most popular RPM-based distributions, and Caixa M\'agica from
Portugal who is well-known in Portuguese-speaking countries. This
distribution is again RPM-based, and also upstream author of the
\texttt{apt RPM} tool. For the successor project Mancoosi (for
\emph{Managing the Complexity of the Open Source Infrastructure})
Pixart from Argentina joined in with its Debian-based distribution.
EDOS started in October 2004 and ended in June 2007. Mancoosi started
in February 2008 for a duration of 3 years.

The EDOS project was relatively broad in scope and had workpackages 
on the following subjects:
\begin{itemize}
\item formal management of software dependencies
\item flexible testing framework 
\item peer-to-peer content dissemination system
\item metrics and evaluation
\end{itemize}

We will in this paper let the last three of these workpackages aside
since the authors haven't been involved in these, and present from
EDOS only the workpackage on dependency management. We decided to
focus on the problem of distribution coherence from the release
manager's point of view, and therein on one basic question: Is it
possible, for a given user selection of packages, to install these
when only the packages from this repository are available? We were
only taking into account package relationships that are expressed by
the metadata of packages (that is in Debian: the \path{control} file).
Relevant results and applications for Debian will be presented in
Section~\ref{sec:edos}.

The successor project Mancoosi again has several workpackages. The
stream on dependency management takes off where EDOS has ended and
tries to extend our previous results to build better tools for the
system administrator who wants to perform a system upgrade or package
installation on a real system. More about this will be discussed in
Section~\ref{sec:mancoosi}.

EDOS has developed its own terminology which Mancoosi continues to
use:
\begin{description}
\item[Installer] A tool to unpack and configure, upgrade, or remove a
 locally available package on a local system. In Debian:
 \texttt{dpkg}.
\item[Meta-Installer] A tool to resolve (higher level) user requests
 of installing, upgrading, or removing packages on a system. This tool
 will have to access possibly remote packages repositories, and
 construct a sequence of commands for an installer. In Debian:
 \texttt{apt-get}, \texttt{aptitude}, \texttt{dselect}.
\item[Metadata] of a package is the data that can be statically (that
 is, without performing an actual installation) extracted from a
 package. In case of Debian this is the contents of a packages
 \texttt{control} file, which flows into APT package lists
 (\texttt{Packages} and \texttt{Sources}).
\end{description}

%% file: edos.tex
\section{The Past: EDOS}
\label{sec:edos}

\subsection{Formalization of Inter-Package Relations}
\label{sec:edos-formalisation}

One of the first objectives of the EDOS project was to establish a
simple mathematical model of a (GNU/Linux) distribution. We decided to
restrict ourselves in the context of EDOS to relations between
packages as they are seen by a meta-installer. Though the model is
general enough to describe the essential features of common packaging
systems (in particular Debian and RPM) we will focus in the following on
the modeling of the package relations as found in Debian.

The Debian policy lists different possible relations between binary
packages: Depends, Recommends, Suggests, Pre-Depends, Enhances, and
Conflicts. The Replaces relation concerns only the installer (not the
meta-installer), and the same seems to be true for the Breaks relation
(which wasn't included in policy anyway at the time of the EDOS
project). Relations between source packages and binary packages are
not of interest for us. However, we have to take into account Provides
(that is, virtual packages), and the fact that relations may be
disjunctive (e.g., \verb+a|b|c+), and may be qualified by constraints
on version numbers.

We decided to ignore relations that are not essential for a
meta-installer in order to decide about installability. This
eliminates Suggests and Enhances from our list of interesting
relations, and we also decided to ignore Recommends relations.
Pre-Depends can for our purposes be identified with Depends.

\begin{figure}
  \begin{center}
  {\tt
  \begin{tabular}{ll@{\qquad}|@{\qquad}ll}
    Package:& a              & Package:& a\\
    Version:& 1              & Version:& 1\\
    Depends:& b, c|d($>$=2) &
    Depends:& b(=2)|b(=3),\\
    &&&c(=3)|d(=2)|d(=3)
    \\[2ex]

    Package:& b              & Package:& b\\
    Version:& 2              & Version:& 2
    \\[2ex]
    Package:& b              & Package:& b\\
    Version:& 3              & Version:& 3
    \\[2ex]
    Package:& c              & Package:& c\\
    Version:& 3              & Version:& 3  \\
    Conflicts:& b            & Conflicts:& b(=2),b(=3)
    \\[2ex]
    Package:& d              & Package:& d\\
    Version:& 1            & Version:& 1
    \\[2ex]
    Package:& d              & Package:& d\\
    Version:& 2            & Version:& 2
    \\[2ex]
    Package:& d              & Package:&d\\
    Version:& 3            & Version:& 3
  \end{tabular}}
\end{center}
  \caption{\label{fig:expansion}A distribution (to the left) and its
  expansion (to the right).}
\end{figure}

This leaves us with Depends and Conflicts. The next question was how
to handle constraints on version numbers like \verb|>= 1:2.3.4-5|. We
decided to not complicate our model with version numbers and their
comparison, and to expand version constraints: given a package in a
package dependency we replace it by the disjunction of all versions of
that package that exist in the current distribution. In case of a
conflict we replace the package by the set of all versions of that
package.  An example of that expansion is given in
Figure~\ref{fig:expansion}.

This expansion has the advantage that we get rid of constraints on
version numbers, but it has the drawback that this expansion is
always relative to a set of available packages. This might pose a
problem when one wants to make the expansion incremental. For
instance, if the original distribution is extended by a new version 4
of package \texttt{d} we would have to reconsider in the expansion all
packages that have a relation to \texttt{d}. In our example, that
means that we have to change the Depends line of package \texttt{a} and
add \texttt{|d(=4)}. 

\begin{figure}
  \begin{center}
    {\tt
    \begin{tabular}{ll@{\qquad}|@{\qquad}ll}
    Package:& a              & Package:& a\\
    Provides:& v\\
    &                        & Package:& b\\
    Package:& b              & Depends:& w\\
    Provides:& v\\             
    Depends:& w              & Package:& v\\   
    &                        & Depends:& a|b
    \\[2ex]
    Package:& c              & Package:& c\\
    Provides:& w             & Conflicts:& d\\
    Conflicts:& w\\
    &                        & Package:& d\\
    Package:& d              & Conflicts:& c\\
    Provides:& w\\
    Conflicts:& w            & Package:& w\\
    &                        & Depends:& c|d\\
  \end{tabular}}
  \end{center}
  \caption{\label{fig:expansion-virtual}A distribution involving
    virtual packages (to the left) and its expansion (to the
    right). Version numbers are omitted.}
\end{figure}

Expansion also introduces explicitly the virtual package which depends
on all packages that provide it. Special care has to be taken with
conflicts on virtual packages as a package may at the same time
provide a virtual package and conflict with it. Section 7.4 of the
Debian policy states that in this case the package conflicts with each
package providing that virtual package, with the exception that the
package doesn't conflict with itself.  An example of an expansion
involving virtual packages is given in
Figure~\ref{fig:expansion-virtual}.

We can now state the formal definition of a package and a repository:
\begin{definition}
  A \emph{package} is pair consisting of a name and a version number.
\end{definition}
Note that we have not defined what package names and version numbers are,
it suffices for us that we can know when two names or version numbers
are equal (as we assume that we are working with an expanded
repository).

\begin{definition}
  \label{def:repository}
  A \emph{repository} is a tuple $R = (P,D,C)$ where $P$ is a set of
  packages, $D : P \rightarrow \P(\P(P))$ is the
  dependency function (we write $\P(X)$ for the set of
  subsets of $X$), and $C \subseteq P \times P$ is the conflict
  relation.  The repository must satisfy the following conditions:
  \begin{itemize}
  \item The relation $C$ is symmetric, i.e., $(\pi_1,\pi_2) \in C$ if and
    only if $(\pi_2,\pi_1) \in C$ for all $\pi_1,\pi_2 \in P$.
  \item Two packages with the same name but different versions
    conflict, that is, if $\pi_1 = (u,v_1)$ and $\pi_2 = (u,v_2)$ with
    $v_1 \neq v_2$, then $(\pi_1,\pi_2) \in C$.
  \end{itemize}
\end{definition}
In this definition, the function $D$ yields for any package the set of
all its dependencies. All these dependencies must be satisfied
simultaneously. If any such dependency is a set with more than one
element than this set is understood as a set of alternatives. The last
restriction, stating that two different versions of the same package
are in an implicit conflict, is specific to Debian (RPM does note have
this \emph{a priori} restriction).

It is now straightforward to translate an expanded \path{Packages}
file into a repository according to Definition~\ref{def:repository}.
For the expanded \path{Packages} file on the right of
Figure~\ref{fig:expansion}, for example, we obtain $(P,D,C)$ as
follows:
\begin{eqnarray*}
  P & = & \{(a,1),(b,2),(b,3),(c,3),(d,1),(d,2),(d,3)\}\\
  D(a,1) & = & \{\{(b,2),(b,3)\},\{(c,3),(d,2),(d,3)\}\}\\
  D(b,2) & = & \emptyset\\
  &\cdots&\\
  C & = & \{ ((b,2),(b,3)),((b,3),(b,2)),((c,3),(b,2)),((b,2),(c,3)),\ldots \}
\end{eqnarray*}

\begin{definition}
  \label{def:installation}
  An \emph{installation} of a repository $R = (P,D,C)$ is a subset $I$
  of $P$, giving the set of packages installed on a system.  An
  installation is \emph{healthy} when the following conditions hold:
  \begin{itemize}
  \item \textbf{Abundance:} Every package has what it needs.
    Formally, for every $\pi \in I$, and for every dependency $d \in
    D(\pi)$ we have $I \cap d \neq \emptyset$.
    \item \textbf{Peace:} No two packages conflict.  Formally, $(I \times I)
    \cap C = \emptyset$.
  \end{itemize}
\end{definition}

\begin{definition}
  A package $\pi$ of a repository~$R$ is \emph{installable} if there
  exists a healthy installation~$I$ such that $\pi \in I$.  Similarly,
  a set of packages $\Pi$ of $R$ is \emph{co-installable} if there
  exists a healthy installation $I$ such that $\Pi \subseteq I$.
\end{definition}

Note that because of conflicts, every member of a set $X \subseteq P$
may be installable without the set $X$ being co-installable. One can
even show that not co-installable sets of minimal size can be
arbitrary large: Let, for a given number $n$, $R_n$ be the following
repository:
\begin{eqnarray*}
  P & = & \{a_1,\ldots,a_n,b_1,\ldots,b_n\}\\
  D(a_i) & = & \{\{b_1,\ldots,b_{i-1},b_{i+1},\ldots,b_n\}\}\\
  D(b_i) & = & \emptyset \\
  C & = & \{ (b_i,b_j) \mid i \neq j \}
\end{eqnarray*}
In this repository, every package $a_i$ depends on the disjunction of
all packages $b_j$ with $j\neq i$. Hence, any incomplete collection of
packages $a$ is co-installable: if package $a_i$ is a package missing
from that collection then we can simply satisfy all dependencies by
installing package $b_i$. Installing all packages $a$ together,
however, would require to install at least two different packages $b$.
Since any two different packages $b$ are in conflict this is not possible.

The desirable property that we want to ensure for a repository $R$ is
the following:

\begin{definition}
  A repository $R$ is {\em trimmed} if every package $\pi \in R$ is
  installable with respect to $R$ itself.
\end{definition}

In Debian lingo this translates to the fact that no package in the
repository is ``broken'', i.e. that there is at least one possible
installation in which any given package is installable. If this is not
the case then that particular Debian distribution will be shipping
packages that users will never be able to install.

\subsection{Results, Tools, and Applications}

\subsubsection{Result: Installability is NP-complete}
\label{sec:npcomplete}
Based on the formalization given in
Section~\ref{sec:edos-formalisation} one can now quite easily show
that the problem whether a given package is installable in a given
repository is logarithmic-space equivalent to the famous SAT problem.
This means two things:
\begin{enumerate}
\item One can construct for any installability problem a SAT problem
  such that the former has a solution if and only the latter has a
  solution \cite{edos-wp2d1,edos2006ase}.
\item One can construct for any SAT problem an installability problem
  such that the former has a solution if and only the latter has a
  solution \cite{edos-wp2d2}.
\end{enumerate}
The ``logarithmic space'' qualifier means that the construction can be
done with auxiliary memory of size logarithmic in the size of the
given problem. This is necessary to transfer complexity results from
one problem to the other.

For instance, in order to translate an installability problem into a
SAT problem we will interpret a package \texttt{p} as a Boolean
variable with the intuitive meaning that package \texttt{p} is installed in
the chosen solution. Dependencies are translated as implications: If
package \texttt{p} depends on \verb+a,b,c|d,e|f+ (which would be
written $D(p)=\{a,b,\{c,d\},\{e,f\}\}$ according to
Definition~\ref{def:repository}) then this translates to the Boolean
implication:
\[
p \rightarrow \big(a \wedge b \wedge (c \vee d) \wedge (e \vee f)\big)
\]
A conflict, say between packages \texttt{a} and \texttt{b}, is expressed
as the formula $\neg(a \wedge b)$. The formula $p$
expresses that the package $p$ has to installed.  This encoding opens
the way to using existing SAT solving techniques to the resolution of
installability problems (see Section~\ref{sec:edos-debcheck}). Since
one has reductions in both directions one obtains an exact worst-case
complexity:
\begin{theorem}
  The problem whether a given package is installable in a repository is
  NP-complete.
\end{theorem}
On a theoretical level this means that checking installability is
infeasible \emph{in its full generality}. In practice it means as
little as that it is a challenging problem since in practice one does
not encounter randomly chosen repositories.  The repositories we
encounter in reality have a quite particular structure. For instance
we will certainly have few packages with a very high number of reverse
dependencies, and a large number with very few reverse dependencies.
Indeed, the implementation developed in the EDOS project is surprisingly
efficient (see Section~\ref{sec:edos-debcheck}). This apparent
contradiction between theoretical very bad \emph{worst-case} complexity
on the one hand and the existence of implementations that are
surprisingly fast for \emph{selected problem instances} is quite
common in computer science.

\subsubsection{Tools: edos-debcheck, pkglab and ceve}
\label{sec:edos-debcheck}

The {\tt edos-debcheck} utility (available in Debian in the package of
the same name) takes as input a package repository and checks whether
one, several or all packages in the repository are installable with
respect to that repository. This utility is based on the SAT encoding
mentioned in Section~\ref{sec:npcomplete} and employs a
customized Davis-Putnam SAT solver \cite{eenss03}.  Since all
computations are performed in-memory and some of the encoding work is
shared between all packages considered this is significantly faster
than constructing a separate SAT encoding for the installability of
each package, and then running an off-the-shelf SAT solver on it. For
instance, checking installability of all packages of main
testing/amd64 takes only 5 seconds on a dual-core amd64 (emitted
warnings about bad package version numbers and other irregularities are
omitted):
\begin{verbatim}
edos-debcheck </var/lib/apt/lists/..._main_binary-amd64_Packages >out
Parsing package file...  1.2 seconds   21617 packages
Generating constraints...  2.3 seconds
Checking packages... 1.5 seconds
4.692u 0.324s 0:05.03 99.6%	0+0k 0+0io 0pf+0w
\end{verbatim}
An explanation in case of non-installability is given, see
Figure~\ref{fig:uninstallable-detail} for an example.  We have also
developed an RPM version of this tool called \texttt{edos-rpmcheck}.

\texttt{pkglab} is an interpreter for a query language that combines
basic queries to edos-debcheck, resp.\ edos-rpmcheck, with a
functional language which allows to use constructions like
\texttt{map} to manipulate conveniently lists of packages. The
interpreter allows to assign intermediate results to variables.  We
are planning for the future a major overhaul of the query language
with the goal of making it more useful as a scripting language for
applications like the one described in Section~\ref{sec:overwrite}.
The interpreter can load repositories that have been pre-processed by
the \texttt{ceve} parser which can parse and analyze both Debian and
RPM repositories. The Debian package for \texttt{pkglab} is pending
while the \texttt{ceve} package is currently available in experimental.

\subsubsection{Application: Finding Uninstallable Packages in Debian}
\label{sec:edos.debian.net}
\texttt{edos-debcheck} is currently used to monitor the state of
Debian's distributions (\textit{unstable}, \textit{testing},
\textit{stable}), as well as Skolelinux and Debian GNU/kFreeBSD.  The
results of the analysis are available at
\url{http://edos.debian.net/edos-debcheck}.

There are different reasons why non-installable packages actually
exist in these distributions. One important reason is that most of the
binary packages are architecture dependent, that is there is one
package per architecture. As a consequence, when accessing the reasons
for non-installability of packages we have to take into account all
possible Debian architectures.

The meta-data of a binary package are generated during the package
compilation from the meta-data in the source package, and may depend
on the actual compilation environment or conditional code in the
source package. As a consequence, the metadata of a package with the
same package name and version may vary from architecture to
architecture.

\begin{itemize}
\item The \textit{unstable} distribution is in fact the staging ground
  for building releasable distributions. Packages that depend on each
  other enter this distribution in an arbitrary order which depends on
  when a developer uploads a package, or on when a package is compiled
  and uploaded by an autobuilder (these are daemons that compile
  packages for the various architectures). For instance, package $a$
  may depend on package $b$, and the developer of $a$ uploads a
  package for the architecture \texttt{i386} while the developer of
  $b$ uploads his package for \texttt{amd64} (he should have tested
  package $b$ using a locally built binary package of $a$ on
  \texttt{amd64}). In this case, $a$ is uninstallable in the
  repository for \texttt{i386} until the \texttt{i386} autobuilder
  daemon uploads the binary package for $b$. This is illustrated by
  Figure~\ref{fig:uninstallable-sid}, the numbers of uninstallable
  packages in sid are indeed varying from day to day.

  As a consequence, transient non-installability errors are normal in
  the \emph{unstable} distribution. Persistent errors, however, indicate 
  a potential problem.
\item A package $a$ may depend on package $b$, but $b$ is not
  available on all architectures $a$ is available on. This may be due
  to the fact that there is a problem with compiling $b$ on some
  architectures, or that $a$ has a too liberal architecture
  specification.
\item A special case of the latter is that $a$ has its architecture
  set to \texttt{all}. This indicates a binary package that is in fact
  the same on all architectures, and hence exists only once in the
  package pool. Package $a$ may, however, depend on a package $b$
  which is architecture \emph{dependant} but does not exist for every
  architecture.  Introducing a field ``Installs-to'' in the syntax of
  control files (as proposed in Bug report
  \#436733\footnote{\url{http://bugs.debian.org/436733}}) would allow
  to fix this.

  Packages which aren't installable on any of the architectures of a
  distribution are more likely due to an error. This may happen with
  packages that are installable in some architecture that has been
  part of a distribution in the past, but which has been removed since
  then. Another possible reason is dependency on a package that had to
  be removed from a distribution, for instance due to licensing problems
  or grave bugs.
\end{itemize}	

\begin{sidewaysfigure}
unstable/main:
\[
\begin{array}{|l|l|l|l|l|l|l|l|l|l|l|l|l|l|l|l|l|l|l|}
\hline
\textrm{Date}&\textrm{alpha}&\textrm{amd64}&\textrm{arm}&\textrm{armel}&
\textrm{hppa}&\textrm{hurd-i386}&\textrm{i386}&\textrm{ia64}&\textrm{m68k}&
\ldots&\textbf{some}&\textbf{every}\\
\hline
22/06
&	949 (325)
& 	121 (80)
& 	604 (126)
& 	609 (103)
& 	613 (132)
& 	4445 (1333)
& 	228 (131)
& 	456 (120)
& 	8943 (4583)
&	\ldots
& 	10222 (5163)
& 	41 (12)\\
\hline
\Delta
& 	\inred{+20\mathord{/-}2}
& 	\ingreen{+7\mathord{/-}11}
& 	\ingreen{+22\mathord{/-}24}
& 	\ingreen{+28\mathord{/-}81}
& 	\ingreen{+24\mathord{/-}34}
&	\ingreen{+10\mathord{/-}38}
& 	\inred{+31\mathord{/-}7}
& 	\inred{+26\mathord{/-}21}
& 	\inred{+21\mathord{/-}10}
&	\ldots
& 	\inred{+44\mathord{/-}5}
& 	\ingreen{+0\mathord{/-}7}\\
\hline
21/06
& 	931 (312)
& 	125 (78)
& 	606 (132)
& 	662 (117)
& 	623 (141)
& 	4473 (1339)
& 	204 (109)
& 	451 (121)
& 	8932 (4586)
&	\ldots
& 	10183(5141)
& 	48(12)\\
\hline
\Delta
& 	\inred{+44\mathord{/-}0}
& 	+1\mathord{/-}1
& 	\inred{+18\mathord{/-}7}
& 	\inred{+52\mathord{/-}12}
& 	\inred{+84\mathord{/-}0}
& 	\inred{+44\mathord{/-}2}
& 	\inred{+56\mathord{/-}0}
&	\inred{+58\mathord{/-}0}
& 	\inred{+34\mathord{/-}5}
&	\ldots
& 	\ingreen{+13\mathord{/-}22}
& 	\ingreen{+0\mathord{/-}1}\\
\hline
20/06
& 	887(287)
& 	125(78)
& 	595(121)
& 	622(108)
& 	539(112)
& 	4431(1337)
& 	148(92)
& 	393(103)
& 	8903(4585)
&	\ldots
& 	10192(5150)
& 	49(13)\\
\hline
\Delta
& 	 \inred{+90\mathord{/-}5}
& 	 \ingreen{+6\mathord{/-}65}
& 	 \ingreen{+17\mathord{/-}77}
& 	 \inred{+21\mathord{/-}14}
& 	 \ingreen{+14\mathord{/-}63}
& 	 \inred{+15\mathord{/-}2}
& 	 \ingreen{+19\mathord{/-}65}
& 	 \ingreen{+13\mathord{/-}64}
& 	 \inred{+26\mathord{/-}15}
&	 \ldots
& 	 \inred{+28\mathord{/-}9}
& 	 \ingreen{+1\mathord{/-}2}\\
\hline
19/06
& 	802(273)
& 	184(83)
& 	655(129)
& 	615(109)
& 	588(113)
&	4418(1338)
&	194(94)
&	444(107)
&	8892(4583)
&	\ldots
&	10173(5148)
&	50(13)\\
\hline
\Delta
&	\inred{+6\mathord{/-}0}
&	\ingreen{+2\mathord{/-}7}
&	\ingreen{+2\mathord{/-}113}
&	\ingreen{+1\mathord{/-}8}
&	\ingreen{+5\mathord{/-}18}
&	\ingreen{+2\mathord{/-}221}
&	+3\mathord{/-}3
&	\ingreen{+5\mathord{/-}7}
&	\ingreen{+1\mathord{/-}37}
&	\ldots
&	\ingreen{+1\mathord{/-}207}
&	\inred{+1\mathord{/-}0}\\
\hline
18/06
&	796(270)
&	189(87)
&	766(145)
&	622(114)
&	601(120)
&	4637(1380)
&	194(96)
&	446(109)
&	8928(4588)
&	\ldots
&	10379(5187)
&	49(13)\\
\hline
\Delta
&	\inred{+5\mathord{/-}0}
&	\ingreen{+4\mathord{/-}8}
&	\inred{+115\mathord{/-}76}
&	\ingreen{+5\mathord{/-}64}
&	\ingreen{+0\mathord{/-}21}
&	\inred{+6\mathord{/-}3}
&	\inred{+4\mathord{/-}1}
&	\ingreen{+1\mathord{/-}76}
&	+5\mathord{/-}5
&	\ldots
&	\inred{+25\mathord{/-}2}
&	+0\mathord{/-}0\\
\hline
17/06
&	791(268)
&	193(92)
&	727(157)
&	681(142)
&	622(132)
&	4634(1379)
&	191(93)
&	521(132)
&	8928(4589)
&	\ldots
&	10356(5167)
&	49(13)\\
\hline
\Delta
&	+12\mathord{/-}12
&	\inred{+11\mathord{/-}1}
&	\ingreen{+14\mathord{/-}57}
&	\ingreen{+15\mathord{/-}74}
&	\ingreen{+67\mathord{/-}105}
&	\ingreen{+4\mathord{/-}32}
&	\ingreen{+4\mathord{/-}42}
&	\ingreen{+9\mathord{/-}67}
&	\inred{+16\mathord{/-}1}
&	\ldots
&	\ingreen{+8\mathord{/-}19}
&	\ingreen{+0\mathord{/-}1}\\
\hline
16/06
&	791(263)
&	183(82)
&	770(175)
&	740(154)
&	660(156)
&	4662(1380)
&	229(96)
&	579(145)
&	8913(4575)
&	\ldots
&	10367(5179)
&	50(13)\\
\hline
\end{array}
\]
\caption[Result of edos-debcheck on
  unstable/main]{\label{fig:uninstallable-sid}Summary of results of running
  edos-debcheck on unstable/main between June 16 and June 22,
  2008. The architectures \textit{mips}, \textit{mipsel},
  \textit{powerpc}, \textit{s390}, and \textit{sparc} are omitted from
  this table for lack of space.

  In each day's listing, the first number is the number of non-installable
  packages, while the number in parentheses is the number of
  non-installable packages that are architecture-specific. Lines
  marked $\Delta$ give the number of packages becoming uninstallable
  the following day (+), resp. that are no longer uninstallable
  (-). This field is colored red when the total number of
  uninstallable packages is increasing, green when that number is decreasing.
  
  \medskip
  Results of a current run can be found at \url{http://edos.debian.net/edos-debcheck/unstable.php}.}
\end{sidewaysfigure}

\begin{sidewaysfigure}
testing/main:

\[
\begin{array}{|l|l|l|l|l|l|l|l|l|l|l|l|l|l|l|l|l|l|l|}
\hline
\textrm{Date}&\textrm{alpha}&\textrm{amd64}&\textrm{arm}&\textrm{armel}&
\textrm{hppa}&\textrm{i386}&\textrm{ia64}&\textrm{mips}&\textrm{mipsel}&
\textrm{powerpc}&\textrm{s390}&\textrm{sparc}&\textbf{some}&\textbf{every}\\
\hline
23/06
&	367 (7)
&	14 (2)
&	217 (4)
&	348 (21)
&	369 (9)
&	12 (4)
&	48 (3)
&	267 (3)
&	269 (3)
&	21 (3)
&	56 (3)
&	24 (3)
&	628 (32)
&	8 (2)\\
\hline
\Delta
&	+0\mathord{/-}0
&	+0\mathord{/-}0
&	\ingreen{+0\mathord{/-}1}
&	+0\mathord{/-}0
&	+0\mathord{/-}0
&	+0\mathord{/-}0
&	+0\mathord{/-}0
&	+0\mathord{/-}0
&	+0\mathord{/-}0
&	\ingreen{+0\mathord{/-}3}
&	+0\mathord{/-}0
&	+0\mathord{/-}0
&	+0\mathord{/-}0
&	+0\mathord{/-}0\\
\hline
22/06
&	367 (7)
&	14 (2)
&	218 (4)
&	348 (21)
&	369 (9)
&	12 (4)
&	48 (3)
&	267 (3)
&	269 (3)
&	24 (4)
&	56 (3)
&	24 (3)
&	628 (32)
&	8 (2)\\
\hline
\Delta
&	+0\mathord{/-}0
&	+0\mathord{/-}0
&	+0\mathord{/-}0
&	+0\mathord{/-}0
&	+0\mathord{/-}0
&	+0\mathord{/-}0
&	+0\mathord{/-}0
&	\ingreen{+0\mathord{/-}3}
&	\ingreen{+0\mathord{/-}3}
&	+0\mathord{/-}0
&	\ingreen{+0\mathord{/-}3}
&	\ingreen{+0\mathord{/-}3}
&	+0\mathord{/-}0
&	+0\mathord{/-}0\\
\hline
21/06
&	367 (7)
&	14 (2)
&	218 (4)
&	348 (21)
&	369 (9)
&	12 (4)
&	48 (3)
&	270 (4)
&	272 (4)
&	24 (4)
&	59 (4)
&	27 (4)
&	628 (32)
&	8 (2)\\
\hline
\Delta
&	+0\mathord{/-}0
&	\ingreen{+0\mathord{/-}3}
&	\ingreen{+0\mathord{/-}3}
&	\ingreen{+0\mathord{/-}9}
&	+0\mathord{/-}0
&	+0\mathord{/-}0
&	+0\mathord{/-}0
&	+0\mathord{/-}0
&	+0\mathord{/-}0
&	+0\mathord{/-}0
&	+0\mathord{/-}0
&	+0\mathord{/-}0
&	+\ingreen{0\mathord{/-}7}
&	\ingreen{+0\mathord{/-}3}\\
\hline
20/06
&	367 (7)
&	17 (3)
&	221 (5)
&	357 (24)
&	369 (9)
&	12 (4)
&	48 (3)
&	270 (4)
&	272 (4)
&	24 (4)
&	59 (4)
&	27 (4)
&	635 (35)
&	11 (3)\\
\hline
\Delta
&	\inred{+7\mathord{/-}0}
&	\inred{+3\mathord{/-}0}
&	\inred{+4\mathord{/-}3}
&	\ingreen{+3\mathord{/-}27}
&	\inred{+4\mathord{/-}0}
&	\inred{+3\mathord{/-}0}
&	\inred{+3\mathord{/-}0}
&	\ingreen{+5\mathord{/-}11}
&	\inred{+5\mathord{/-}0}
&	\inred{+5\mathord{/-}0}
&	\inred{+5\mathord{/-}0}
&	\inred{+5\mathord{/-}0}
&	\ingreen{+5\mathord{/-}16}
&	\inred{+3\mathord{/-}0}\\
\hline
19/06
&	360 (5)
&	14 (2)
&	220 (6)
&	381 (31)
&	365 (8)
&	9 (3)
&	45 (2)
&	276 (2)
&	267 (2)
&	19 (2)
&	54 (2)
&	22 (2)
&	646 (42)
&	8 (2)\\
\hline
\Delta
&	+0\mathord{/-}0
&	+0\mathord{/-}0
&	+0\mathord{/-}0
&	+0\mathord{/-}0
&	+0\mathord{/-}0
&	+0\mathord{/-}0
&	+0\mathord{/-}0
&	+0\mathord{/-}0
&	+0\mathord{/-}0
&	+0\mathord{/-}0
&	+0\mathord{/-}0
&	+0\mathord{/-}0
&	+0\mathord{/-}0
&	+0\mathord{/-}0\\
\hline
18/06
&	360 (5)
&	14 (2)
&	220 (6)
&	381 (31)
&	365 (8)
&	9 (3)
&	45 (2)
&	276 (2)
&	267 (2)
&	19 (2)
&	54 (2)
&	22 (2)
&	646 (42)
&	8 (2)\\
\hline
\Delta
&	+0\mathord{/-}0
&	+0\mathord{/-}0
&	+0\mathord{/-}0
&	+0\mathord{/-}0
&	+0\mathord{/-}0
&	+0\mathord{/-}0
&	+0\mathord{/-}0
&	+0\mathord{/-}0
&	+0\mathord{/-}0
&	+0\mathord{/-}0
&	+0\mathord{/-}0
&	+0\mathord{/-}0
&	+0\mathord{/-}0
&	+0\mathord{/-}0\\
\hline
17/06
&	360 (5)
&	14 (2)
&	220 (6)
&	381 (31)
&	365 (8)
&	9 (3)
&	45 (2)
&	276 (2)
&	267 (2)
&	19 (2)
&	54 (2)
&	22 (2)
&	646 (42)
&	8 (2)\\
\hline
\end{array}
\]

stable/main:
\[
\begin{array}{|l|l|l|l|l|l|l|l|l|l|l|l|l|l|l|l|l|l|l|}
\hline
\textrm{Date}&\textrm{alpha}&\textrm{amd64}&\textrm{arm}&
\textrm{hppa}&\textrm{i386}&\textrm{ia64}&\textrm{mips}&\textrm{mipsel}&
\textrm{powerpc}&\textrm{s390}&\textrm{sparc}&\textbf{some}&\textbf{every}\\
\hline
23/06
& 	184 (0)
& 	13 (0)
& 	96 (2)
& 	189 (0)
& 	0 (0)
& 	67 (0)
& 	185 (0)
& 	186 (0)
& 	13 (0)
& 	183 (0)
& 	144 (4)
& 	235 (6)
& 	0 (0)\\
\hline
\end{array}
\]
\caption[Result of edos-debcheck on testing/main and
  stable/main]{\label{fig:uninstallable-testing-satble}The same
  statistics as in Figure~\ref{fig:uninstallable-sid} now for testing
  and stable (only one day shown since no variation).}
\end{sidewaysfigure}

\begin{figure}
  \begin{tabularx}{\linewidth}{|l|l|l|X|}
    \hline
    Package & Since & Version & Explanation\\
    \hline
    $\ldots$ & $\ldots$ & $\ldots$ & $\ldots$\\
    \hline
    \textit{calendarserver}
    & 20 Jun 08
    & 1.2.dfsg-3
    & \texttt{calendarserver (= 1.2.dfsg-3) depends on
      python-twisted-calendarserver (>= 0.2.0.svn19773-3) \{NOT AVAILABLE\}}\\
    \hline
    \textit{camping}
    & 21 Jun 08
    & 1.5+svn242-1
    & \texttt{camping (= 1.5+svn242-1) depends on rails \{rails (= 2.0.2-2)\}
      rails (= 2.0.2-2) depends on rdoc (>> 1.8.2) \{rdoc (= 4.2)\} rdoc (= 4.2)
      depends on rdoc1.8 \{rdoc1.8 (= 1.8.7.22-1)\}} \\
    \hline
    $\ldots$ & $\ldots$ & $\ldots$ & $\ldots$\\
    \hline
    \textit{rdoc1.8}
    & 21 Jun 08
    & 1.8.7.22-1
    & \texttt{rdoc1.8 (= 1.8.7.22-1) depends on ruby1.8 (>= 1.8.7.22-1)
      \{NOT AVAILABLE\}}\\
    \hline
    $\ldots$ & $\ldots$ & $\ldots$ & $\ldots$\\
    \hline
    shoes
    & 21 Jun 08
    & 0.r396-4
    & \texttt{shoes (= 0.r396-4) depends on libgems-ruby1.8
      \{libgems-ruby1.8 (= 1.1.1-1)\} libgems-ruby1.8 (= 1.1.1-1) depends
      on rdoc1.8 \{rdoc1.8 (= 1.8.7.22-1)\}}\\
    \hline
  \end{tabularx}
  \caption[Listing of uninstallable packages with
    explanation]{\label{fig:uninstallable-detail}An excerpt from the
    list of uninstallable packages in sid/i386 main for June 22,
    2008. In the explanation field, available versions of a package
    are indicated between curly brackets. Lines may refer to packages
    shown non-installable elsewhere, like the packages
    \texttt{camping} and \texttt{shoes} being not-installable because
    it need \texttt{rdoc1.8}. Package names written in
    \textit{italics} in the left column have Architecture=all.

    \medskip
    Results of a current run can be found at
    \url{http://edos.debian.net/edos-debcheck/results/unstable/latest/i386/list.php}.}
\end{figure}

\subsubsection{Application: Debian Weather}

\begin{figure}
  \def\wpic#1{\includegraphics[height=1cm]{pics/weather-#1_sml}}
    \begin{tabular}{lcccccccccccccccc}
      \multicolumn{5}{l}{Stable:}\\
      \wpic{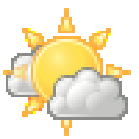} & \wpic{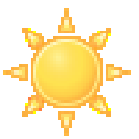} & \wpic{clear} &
       \wpic{few-clouds} & \wpic{clear} & \wpic{clear} &
       \wpic{few-clouds} & \wpic{few-clouds} & \wpic{clear}\\
      \multicolumn{5}{l}{Testing:}\\
      \wpic{few-clouds} & \wpic{clear} & \wpic{few-clouds} &
       \wpic{few-clouds} & \wpic{clear} & \wpic{clear} &
       \wpic{few-clouds} & \wpic{few-clouds} & \wpic{clear}\\
      \multicolumn{5}{l}{Unstable:}\\
      \wpic{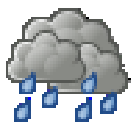} & \wpic{clear} & \wpic{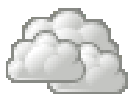} &
       \wpic{overcast} & \wpic{clear} & \wpic{few-clouds} &
       \wpic{few-clouds} & \wpic{few-clouds} & \wpic{clear}\\
      alpha & amd64 & arm & hppa & i386 & ia64 & mips & mipsel & powerpc
    \end{tabular}
  \caption{\label{fig:weather}The Debian weather for June 27, 2008:
    Mostly sunny in stable and testing, at places overcast and rainy
    in unstable.}
\end{figure}

This is more of a fun application. Based on the numbers of the tool
described in Section~\ref{sec:edos.debian.net} a ``weather report'' of
Debian is generated which indicates the percentage of non-installable
packages for the different distributions and architectures. The
interpretation is as follows:
\[
\begin{array}{l|l}
  \textrm{clear}      &	< 1\%\\
  \hline
  \textrm{few clouds} & 1\% \ldots 2\%\\
  \hline
  \textrm{clouds}     & 2\% \ldots 3\%\\
  \hline
  \textrm{showers}    & 3\% \ldots 4\%\\
  \hline
  \textrm{storm}      &	> 4\%
\end{array}
\]
An example weather report is given in
Figure~\ref{fig:weather}. Applets for Gnome and KDE are available.

The daily updated Debian weather is available on the web at
\url{http://edos.debian.net/weather}.

\subsubsection{Application: Finding File Conflicts in Debian}
\label{sec:overwrite}
A Debian installation has the concept of files owned by packages. If
one tries to install a new package that would hijack a file owned by
another package this will make (with some exceptions, see below) the
installation fail, like this:

\begin{small}
\begin{verbatim}
Unpacking gcc-avr (from .../gcc-avr_1%3a4.3.0-1_amd64.deb) ...
dpkg: error processing /var/cache/apt/archives/gcc-avr_1%3a4.3.0-1_amd64.deb
 (--unpack):
 trying to overwrite `/usr/lib64/libiberty.a', which is also in package
 binutils
dpkg-deb: subprocess paste killed by signal (Broken pipe)
Errors were encountered while processing:
 /var/cache/apt/archives/gcc-avr_1%3a4.3.0-1_amd64.deb
E: Sub-process /usr/bin/dpkg returned an error code (1)
\end{verbatim}
\end{small}

Our aim is to detect these errors by analyzing the Debian
distribution, hopefully before they actually occur on a user machine.

An obvious na{\"\i}ve solution would be to try to install together all
pairs of packages that occur in the distribution. Debian amd64/testing
has currently about 21.000 packages, that would make about 200.000.000
pairs of packages to test, which clearly is not feasible.

A first idea towards a better solution is to only consider those pairs
of packages that actually share at least one file. Luckily, the
information which package contains which file is available in the file
\texttt{Contents} of the distribution. This file contains stanzas like
\begin{verbatim}
...
bin/fbset                       admin/fbset
bin/fgconsole                   utils/console-tools,utils/kbd
...
etc/default/nvidia-kernel       contrib/x11/nvidia-kernel-common
...
\end{verbatim}
In this file, information is indexed by path names of the files
(omitting the initial slash). For every file a comma separated list of
packages containing that file is given where packages are indicated
with their section (a classification of packages by type, like
\texttt{games} or \texttt{admin}), and probably the component if it is
different from \texttt{main} (which can currently be \texttt{contrib}
or \texttt{non-free}). For instance, the file \texttt{/bin/fgconsole}
is provided by the packages \texttt{console-tools} and \texttt{kbd}
which both are in section \texttt{utils}. In fact the
\texttt{Contents} file that can be found on a Debian mirror may be
slightly out of date as this file is generated only once per week.

The \texttt{Contents} file of amd64/testing (as of May 2008) contains
about 2.300.000 entries. It is a trivial programming exercise to
compute from this file a list of pairs of packages that share at least
one file.

Sharing a file does not necessarily mean a bug. There a several
reasons why it may be OK for two packages, say \texttt{A} and
\texttt{B}, to share a file, say {F}:
\begin{enumerate}
\item The two packages are not co-installable by the package
  relationships declared in their distribution, in the sense of
  Section~\ref{sec:edos-formalisation}.
\item One of the packages, say \texttt{A}, declares that it has the
  right to replace files owned by \texttt{B}, by having in its control
  file a stanza \texttt{Replaces: B}.
\item One of the packages, say \texttt{B}, \emph{diverts} the file
  \texttt{F} that it shares with package \texttt{A}. This means that
  if package \texttt{B} is being installed on a system already containing
  package \texttt{A} then \texttt{A}'s version of file {F} will be renamed;
  file \texttt{F} will be restored to its original name when package
  \texttt{B} will be removed. File diversions are declared by invoking
  the tool \texttt{dpkg-divert} from a maintainer script which will
  simply register the diversion request in a system-wide
  database. This database is consulted by \texttt{dpkg} when
  installing files.
  Diversions are not declared in the package control file.
\end{enumerate}

We proceed in two stages in order to find the actual file overwrite
problems:
\begin{enumerate}
\item Co-installability is checked with the \texttt{pkglab} tool
  (see Section~\ref{sec:edos-debcheck}). This is the only tool that can detect
  ``deep'' conflicts between packages. This first phase gives us a
  reduced list of pairs of packages.
\item Knowing which files are diverted by a package poses different
  problems: diversions are registered by the so-called postinst script
  of a package, one of the maintainer scripts that are executed during
  installation (or upgrade, or removal) of a package. This leads to two
  problems:
  \begin{enumerate}
  \item Execution of the postinst script depends on the current state
    of the system, and can in general not be described by a simple list
    of files.
  \item The postinst script is written in a Turing complete language
    (usually Posix shell or bash), which means that exact semantic
    properties are undecidable.
  \end{enumerate}
  
  For this reason, we try in the second phase to install each of the
  pairs of packages remaining after the first phase in a chroot, using
  \texttt{apt-get install}. We then search the install log for file
  overwrite errors.
\end{enumerate}

The following statistics is from the first run performed on April 16th,
2008, on amd64/sid:

\begin{center}
  \begin{tabular}{|l|r|}
    \hline
    Theoretical pairs of packages according to the distribution& 200.000.000 \\
    \hline Pairs of packages sharing a file according to \texttt{Contents}&
    867   \\
    \hline Co-installable pairs among these according to
    \texttt{pkglab} & 102 \\
    \hline File overwrites detected & 27\\
    \hline
  \end{tabular}
\end{center}

Checking co-installability with EDOS pkglab took 30 minutes and gave a
88\% reduction of the search space. Testing the installation of the
remaining 102 pairs of packages still took 2.5 hours. This measures
where taken with a dual-core amd64 at 1.6GHz, using a local Debian
mirror access over a fast LAN.

Detected bugs are tracked in the Debian bug tracking system, and
marked there with user \verb|treinen@debian.org| and usertag
\verb|edos-file-overwrite|.

%% file: mancoosi.tex
\section{Present and Future: Mancoosi}
\label{sec:mancoosi}

\subsection{An Overview of the Mancoosi Project}

Mancoosi picks up the baton from where EDOS left it. So, where to go
from EDOS? Even though some of the theoretical achievements of EDOS
still have some way to go before reaching the practice of all
distributions (including Debian), adoption of EDOS results is ongoing
and is actually extending past the distribution universe; a noteworthy
example is the Eclipse platform, which is moving to SAT solving to
solve inter-plugin dependencies.

On the contrary, one side of the complexity issues introduced by the
overwhelming amount of packages in GNU/Linux distributions has been
neglected by EDOS and is still in need of both research and tool development:
 the \emph{user side} of a distribution.
While EDOS has focused on the \emph{distribution editor side} (i.e. on
who is actually creating the distributions), Mancoosi focuses on who
is actually using a distribution, in particular \emph{system
administrators}.

It is well-known that distributions raise difficult problems for
administrators. Distributions evolve rapidly by integrating new
versions of software packages that are independently developed. System
upgrades may proceed on different paths depending on the current state
of a system and the available software packages, and system
administrators are faced with choices of upgrade paths, and possibly
with failing upgrades. All together, these intertwined problems are
referred to as the \emph{upgrade problem}. The Mancoosi project aims
at developing tools for the system administrator that address the
upgrade problem.

What does constitute an upgrade problem from the point of view of a
system administrator? Intuitively, any possible change to the database
of locally installed packages constitutes an upgrade problem. Such
changes are usually requested to a meta-installer and are well-known
to any system-administrator. Some examples:

\begin{itemize}
 \item \texttt{apt-get install wesnoth}
 \item \texttt{aptitude upgrade cappuccino}
 \item \texttt{apt-get dist-upgrade}
 \item \texttt{aptitude purge emacs22}
 \item \texttt{wajig install vim-full}
\end{itemize}

Each of the above examples poses a simple upgrade problem. Way more
complex upgrade problems can be formed by combining simpler
problems (e.g. posing all the above requests together to a single
meta-installer). Yet more complex problem can be created by exploiting
meta-installer specific features such as requiring specific package
versions or origin suites (think at \texttt{apt} pinning).

A basic principle of the Mancoosi project was that the upgrade
process can be decomposed into two parts: dependency resolution
and upgrade deployment.
While dependency resolution can be thought of  as a static phase, where
without altering the package database a meta-installer has to figure
out if and how to implement the user request, upgrade deployment is more
dynamic and consists of several sub-activities: package download,
package unpacking, maintainer scripts execution \ldots

According to this distinction, the two main avenues pursued by
Mancoosi are:

\begin{description}

 \item[rollback support] Upgrade deployment can fail for various
   reasons easily encountered in system administrator nightmares
   (disks running out of space, 404 while downloading a package,
   maintainer script failures, file overwrites among unrelated
   packages, \ldots).  Depending on how bad the error is, a common
   attempted solution is that of \emph{rolling back} the system,
   partially or completely, to a safe state which predates the upgrade
   attempt. Unfortunately, support for upgrade attempt rollback is
   basically inexistent in state of the art installers. Note that the
   need for a rollback may also occur some time after an upgrade (even
   days or weeks), and
   that in that case one only wants to undo the package upgrade but not
   any other system changes that have been applied in the
   meantime. This means that we are looking for solutions beyond mere
   file system snapshots.

   Mancoosi aims at developing mechanisms that provide for rollback of
   failed upgrade attempts, allowing the system administrator to revert
   the system to the state before the upgrade. In particular, rollback
   is the topic of Mancoosi work packages 2
   and~3.\footnote{\url{http://www.mancoosi.org/work.html}}
  
 \item[dependency solving] The first part of the upgrade problem is
  implemented by state of the art meta-installers, but each of them has
  deficiencies (e.g. incompleteness: the inability to find an upgrade
  path each time one upgrade path does exists).

  Mancoosi aims at developing better algorithms to plan upgrade paths
  based on various information sources about software packages and on
  optimization criteria. Dependency solving is the topic of Mancoosi
  work packages 4 and~5.

\end{description}

As the authors are only marginally involved with rollback support,
that part of the project will not be discussed any further in this
paper. We will for the rest of this paper concentrate on dependency solving.

\subsection{Dependency solving}

As already mentionend, the overall goal of this part of Mancoosi
is improving dependency solving in state of the art meta-installers,
solving some of their deficiencies. More precisely, Mancoosi plans to
address three requirements which are believed to define the ideal to
which any given meta-installer should tend to: completeness,
optimality, efficiency.

\subsubsection{Completeness} The first of these requirements can be
defined as follows:
\begin{definition}
  A meta-installer is \emph{complete} wrt.\ dependency solving iff for
  each possible upgrade problem which has a solution, the
  meta-installer is able to find such a solution.
\end{definition}

Even though not enough details have been given to fully formalize
completeness in this paper, the intuition should be clear: once the
system administrator poses an upgrade problem to its meta-installer of
choice, the meta-installer tries to solve dependencies to fulfill the
user request to determine which changes should be made to the set of
installed packages. \emph{If} a healthy installation satisfying the
user request does exist, then the meta-installer should be able to
propose it as \emph{a} possible way of fulfilling the user request.

Surprising as it might sound, most state of the art meta-installers
are not complete. For instance, upon receiving a request like
\texttt{install p}, \texttt{apt-get} always tries to install the
latest version of \texttt{p} among those available in the package
universe formed by APT repositories. In case the version requirements
of (latest) \texttt{p} are not satisfiable it might well be that
requirements of (previous) \texttt{p} are indeed satisfiable. In such
and similar cases the user is left with the feeling that there is no
way to satisfy her request, while this is actually not the case: this
is a lack of completeness that should be addressed to improve user
experience with meta-installers.

Note that the given example is just a paradigmatic one, more complex
examples built on top of the limited back-tracking capabilities of
other meta-installers can also be provided \cite{edos-wp2d2} (see also
\url{http://www.mancoosi.org/edos/manager.html} for an analysis of the
situation in the year 2006). The general point stressed here is that
legacy meta-installers which are advertised as \emph{the} tools for
system-administrators to interact with the package database of their
machines should be able to solve dependency problems each time it is
possible to do so.

\subsubsection{Optimality} \label{sec:dep-optimality} Once it can be
taken for granted that any possible solution to a dependency problem
can be found, it is natural to ask \emph{which} among all the possible
solutions has to be preferred over the others.

Note that for any given upgrade problem there are in general
several possible solutions. If you consider again the \texttt{install
  p} request posed to apt-get above, a possible solution for it is to
install the version of \texttt{p} whose dependencies are satisfiable
together with all its (transitive) dependencies and be done with that.
Another valid solution is to install the same set of packages
together with a package \texttt{z} which is completely unrelated to
\texttt{p} and that does not inhibit a healthy installation. Whereas
in these two cases it seems obvious that the former has to be
preferred, in the general case there are non obvious choices to be
made. Anyone who has already been faced with \texttt{aptitude} interactive
solution discrimination knows that: in satisfying dependency problems
coming from user requests, trade-offs have to be made.

In fact, even before discussing \emph{how} the optimal solution has
to be found among all alternative solutions of a given upgrade
problem, there is a need to understand which criteria should be used
to define the optimality of a given solution. At the moment some fixed
criteria which are likely to address most user needs are being
considered; here is a handful of examples:
\begin{itemize}
 \item minimize the amount of extra-packages installed with respect to
  those explicitly mentioned in the user request,
 \item minimize the download size of packages required to deploy the
  upgrade solution,
 \item minimize disk usage after the upgrade (a frequent need for
  Debian-based embedded distributions),
 \item upgrade as many packages as possible to the latest available
  version.
 \item \ldots
\end{itemize}

Of course different optimization criteria can be in conflict one with
another. If on one side this brings the upgrade problem in the
vibrating research field of multi-criteria optimization, it also
raises the issue of which interface should be given to users to
specify their optimization preferences. Moreover, the set of possible
optimization criteria should be open-ended as specific user needs
arise every day: APT pinning is a practical example of user
requests that should be taken into account while choosing an optimal
solution, countless other user-specific requirements can be imagined
(e.g.: when you have a choice among two packages choose the one with
less RC bugs, or even blacklist packages maintained by Random J.
Developer as you don't trust him \ldots). For this reason Mancoosi will
also be developing a cross meta-installer language to specify
optimization criteria with a well-defined semantics, to be used by
system-administrators to specify their preferences.

\subsubsection{Efficiency} Once it is settled what properties we want
from the ability of a meta-installer to solve dependencies
(completeness and optimality), the attention can be turned to how we
would like the given tool to reach a solution \ldots and of course we
want it to be efficient in finding it. Even letting aside the
optimization part, dependency solving is per se a NP-complete problem
(see Section~\ref{sec:npcomplete}) hence we cannot hope for a
definitive algorithm or implementation delivering upgrade problem
solution instantaneously in any given case.

Nevertheless we should strive for the most possible efficiency and in
this respect the EDOS results have been encouraging. Mancoosi will focus
on finding efficient algorithms which not only take into account package
installability ``in the void'' (i.e. in some, not specified a priory,
installation), but rather which address upgrades starting from an
existing user installation.

\subsection{A solver competition}

Promising to find \emph{the} most efficient algorithmic solution to
the upgrade problem, implementing both completeness and optimality in
the setting of the Mancoosi project would have been inconsiderate.
This is why Mancoosi chooses a different path: try increasing
the sensibility of the relevant research communities on the upgrade
problem. Historically, the organization of periodic competitions has
been a training factor in pushing further the state of the art in
algorithms and tools for complex problems such as SAT. Examples like the SAT
competition\footnote{\url{http://www.satcompetition.org/}} and SAT
race\footnote{\url{http://www-sr.informatik.uni-tuebingen.de/sat-race-2008/}}
attract yearly research and practitioners willing to challenge their
tools with competitors to determine which is the ``best'' both in terms
of solver capabilities and in terms of execution speed.

Mancoosi will follow a similar path for the upgrade problem faced
routinely by meta-installers. A competition of dependency solvers will
be organized and is planned to be held in parallel with a research
conference on related fields (SAT-solving, linear optimization,
\ldots). While it is too early to have detailed information on how the
competition will be run and organized, some aspects are already clear.

\paragraph{Upgrade problem database}

To run a solver competition you need a corpus of problems that will be
used to challenge the various competitors. In the Mancoosi case the
corpus will be called UPDB for Upgrade Problem DataBase. The way in
which it will be assembled is different from other
competitions. Instead of creating artificial problems by hand (that
would be not only challenging given the typical size of a distribution
repository, but also bear the risk of creating irrelevant problems)
the corpus will be composed of problems submitted by users who
encountered these.

\begin{figure}[tb!]
 \begin{center}
  \includegraphics[width=0.9\textwidth]{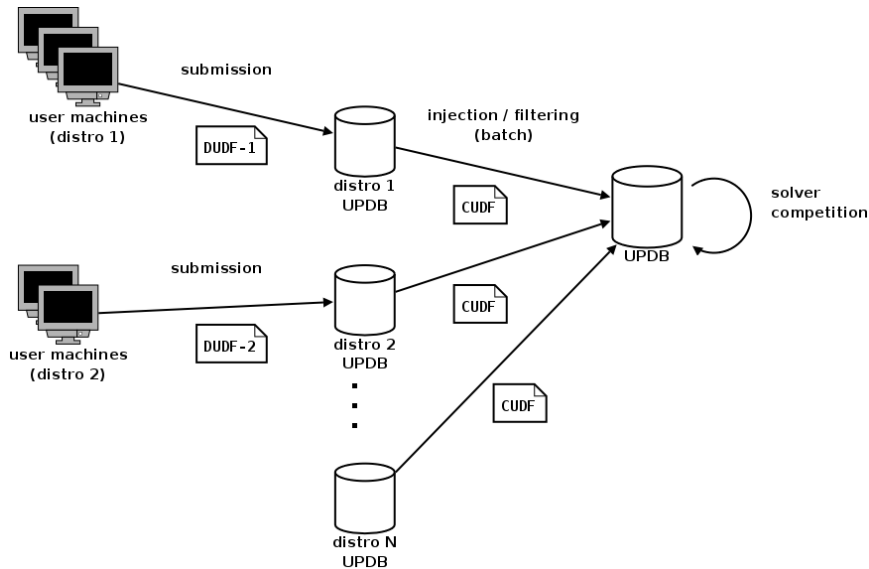}
  \caption{\label{fig:updb-dataflow} Data flow of UPDB submissions,
  from users to the corpus of problems for the competition}
 \end{center}
\end{figure}

All in all, the architecture is similar to that of the Debian
Popularity Contest:\footnote{\url{http://popcon.debian.org/}} users
interested in participating will be asked to install some
special-purpose packages which provide the software to gather data and
submit it to a central repository. In some cases it will probably
be necessary to install modified versions of meta-installers which have been
changed to log enough information to fully describe an upgrade problem.
The architecture of problem submission to UPDB is depicted in
Figure~\ref{fig:updb-dataflow}.

As various distributions are taking part in the Mancoosi
competition, each of them will be providing a staging repository to
which problem submissions will be addressed. One such repository will
be set-up for Debian users as well. As the format of the initial
submission is distribution-specific, a further conversion step into a
common format used to encode problems is needed. Once the conversion
has been done, the upgrade problem is fully abstracted over the origin
distribution and can be fed as input to the various solvers which will
be taking part in the competition.

The Mancoosi project will be both organizing the competition (and this
is the topic of work package 5) and participating in it (work package
4) with a research team which is expert in SAT solving and
optimization techniques and which will be developing ad-hoc algorithms
for the upgrade problem as faced in distributions.

\paragraph{Types of competitions} Different kinds of competitions will
be held. In the beginning it is planned that the optimization criteria
will be fixed and each competitor will specifically be participating
in a selection of them. For example it is likely that we will be having
categories like: no optimization (just solve the upgrade problem no
matter what), minimize the download size of required packages,
minimize disk usage, and so on.

\paragraph{Upgrade Description Formats}

As it can be observed in Figure~\ref{fig:updb-dataflow}, different
format specifications are required before being able to start
collecting upgrade problems from users (that notwithstanding
specification implementations, which will be required as well). Such
specifications are work in progress and are available in the Mancoosi
public repository available at {\small
\url{http://gforge.info.ucl.ac.be/plugins/scmsvn/viewcvs.php/trunk/updb/doc/cudf/?root=mancoosi}}.

The first specification \emph{DUDF (Distribution Upgrade Description
Format)} is meant to describe the format used for the actual
submission of upgrade problems from user machines to the repositories
set up by each distribution interested in collecting upgrade problems.
As the format is in the end distribution-specific, the specifications
describe the overall structure and basic principles of a submission
document, the actual details will be filled in by each distribution
according to the user installers and meta-installers. Interested
distributions are encouraged, once the final version of DUDF will be
ready, to publish notes describing exactly how they are implementing
the distribution-specific part of DUDF.

Roughly, a DUDF document has the following parts:
\begin{enumerate} \setlength{\itemsep}{0mm} \setlength{\parsep}{0mm}
 \item local package status on the user machine
 \item current package universe as known to the meta-installer
 \item requested action
 \item user desiderata (i.e. optimization criteria)
 \item various identifiers (e.g.: distribution identifier, installer
  name and version, meta-installer name and version, \ldots)
 \item outcome of the meta-installer (a new local package status in
  case of success, a failure message otherwise)
\end{enumerate}

A hypothetical (and incomplete) mapping to Debian for the
\texttt{apt-get}, just to give a practical intuition of what can
constitute a DUDF submission, is as follows:
\begin{enumerate} \setlength{\itemsep}{0mm} \setlength{\parsep}{0mm}
 \item \texttt{/var/lib/dpkg/status}
 \item the set of APT binary package lists as stored under
   \texttt{/var/lib/apt/lists/}
 \item the given APT command
 \item current APT pinning settings
 \item ``debian'', ``apt-get'', v\emph{x.y.z}, ``dpkg'', \ldots
 \item ``broken packages, the following packages can not be installed,
  \ldots.''
\end{enumerate}

As sending all the above information can be costly in terms of
submission size, DUDF implements some space-optimizations. The most
important optimization is based on the assumption that most package
lists composing a given package universe are usually only mirrored on
a local machine and are available elsewhere. Hence, by keeping
distribution-specific historical mirrors of a given distribution,
instead of sending whole package lists, a DUDF submission may just
contain package list checksums that can later be looked up in
historical mirrors to recreate the package lists as available on user
machines. In the specific case of Debian, Mancoosi will be keeping
historical mirrors of APT lists for the most widespread
\texttt{apt-get} repositories: not only the official
stable/testing/unstable Debian suites, but also volatile, backports,
debian-multimedia, \ldots

The second, and last, document format involved with the solver
competition is \emph{CUDF (Common Upgrade Description Format)}. That
is the format in which the actual inputs from competition participants
will be encoded in. Contrary to DUDF, CUDF is distribution agnostic
as well as agnostic to any specific installer or meta-installer. A
requirement for any given DUDF document is that it can be converted to
CUDF, during that conversion step all performed space-optimization
will be expanded to obtain a self-contained description of an upgrade
problem.

\subsection{Debian and Mancoosi}

As already mentionend there is no ``official'' relation between the
Mancoosi and Debian projects; however, there are Debian developers in
the ranks of Mancoosi which are interested in giving back to Debian as
much as possible of Mancoosi achievements. This section lists the
foreseeable points of contact between Mancoosi and Debian, it also
points to the available resources for interacting with Mancoosi from
the Debian side.

Probably the main point of interest for Debian in Mancoosi is the
possibility to improve the available algorithms and tools for
dependency solving, both from the point of view of performance and
the point of view of capabilities. To be delivered in Debian, the
possible forthcoming achievements will need cooperation among the
algorithm developers and the developers of meta-installers used in
Debian (apt-get, aptitude, \ldots). The Debian developers involved in
Mancoosi have already taken contact with members of the respective
development teams. Collaborations are needed mainly in two areas:

\begin{description}

 \item[common solver API] It is unlikely that Mancoosi will have the
  energy to port novel dependency resolution algorithms to multiple
  meta-installers, it is more likely that only a proof of concept
  implementation for a single tool will be developed. As Debian is
  also about diversity, it would be preferable to have
  implementations for all the mainstream meta-installers. To this end
  a side-result that will be pursued is the development of a common
  API to let whatever meta-installer interact with an \emph{external
  dependency solver}.  This way it would be possible to develop
  separately meta-installers and plug them into different tools. Such an
  achievement, if reached, would also mean that it will be possible to
  exchange solvers which already exist among different tools, gaining
  flexibility in the overall package manager implementation.

 \item[dependency solving logging] Once the specification of DUDF will
  be finalized, its implementations will basically consist of patches
  (or plugins, where feasible) for meta-installers enabling them to save in
  DUDF format solving attempts originated from upgrade problems. As it
  will be beneficial to have a common format for logging such attempts
  (e.g. for bug reports against apt-get, aptitude, \ldots) we hope to
  spread DUDF implementations in whatever meta-installer is currently
  used in Debian.

\end{description}

On a less implementative side, Mancoosi is welcoming comments from the
Debian community on all aspect of the project. In particular, at the
time of this writing we are interested in comments on what will constitute
\emph{interesting optimization criteria} as those anticipated in
Section~\ref{sec:dep-optimality}. The corpus of collected optimization
criteria is likely to be used as the set of categories to run the
first solver competition. Do not hesitate to get in touch with the
Mancoosi project if you have suggestions on this topic or on anything
else related to the project!

To \textbf{get in touch with Mancoosi} there are various ways.

\begin{itemize}
 \item The official \emph{website} gives general information on the
   Mancoosi project, it is available at \url{http://www.mancoosi.org}
 \item The mailing list to archive public discussions about Mancoosi
   is mancoosi-discuss:
   \url{http://sympa.pps.jussieu.fr/wws/info/mancoosi-discuss}
 \item Then there are also \emph{Debian-specific contacts}
  \begin{itemize}
   \item \url{http://mancoosi.debian.net} has been set-up as a web
     archive of resources for the Debian project offered by Mancoosi.
     At the moment it just contains the historical mirror of APT's
     binary package lists which will be used to implement the
     space-optimization of DUDF.
     
     It also contains an apt-get
     repository of unofficial Debian packages meant as a staging area
     for packages not (yet) accepted in the Debian archive, or simply
     not suitable/interesting enough for it.
   \item the email contact \url{debian@mancoosi.org} is the main
     contact to get in touch with Mancoosi for Debian-related issues,
     questions, comments \ldots Drop a mail to it for more
     information!
  \end{itemize}
\end{itemize}